\documentclass[prl,twocolumn,showpacs,superscriptaddress]{revtex4}

\usepackage{graphicx}
\begin{document}

\title{Effect of boundaries on the force distributions in granular
media}

\author{Jacco H. Snoeijer} \affiliation{Instituut--Lorentz,
Universiteit Leiden, Postbus 9506, 2300 RA Leiden, The Netherlands}

\author{Martin van Hecke} \affiliation{Kamerlingh Onnes Lab,
Universiteit Leiden, Postbus 9504, 2300 RA Leiden, The Netherlands}

\author{Ell\'ak Somfai}\thanks{Present address: Department of Physics, University
of Warwick, Coventry CV4 7AL, U.K.} \affiliation{Instituut--Lorentz,
Universiteit Leiden, Postbus 9506, 2300 RA Leiden, The Netherlands}


\author{Wim van Saarloos} \affiliation{Instituut--Lorentz,
Universiteit Leiden, Postbus 9506, 2300 RA Leiden, The Netherlands}

\date{\today}

\begin{abstract}
The effect of boundaries on the force distributions in granular media
is illustrated by simulations  of 2D packings of frictionless,
Hertzian spheres.  To elucidate  discrepancies between experimental 
observations and theoretical predictions, we distinguish between the 
weight distribution ${\cal P} (w)$ measured in experiments and analyzed 
in  the $q$-model, and the distribution of interparticle forces $P(f)$. 
The latter one is  robust, while ${\cal P}(w)$ can be obtained once the 
local packing geometry and $P(f)$ are known. By manipulating the 
(boundary) geometry, we show that ${\cal P}(w)$ can be varied drastically.



\end{abstract}


\maketitle 

A crucial property of granular materials is their
he\-te\-ro\-ge\-ne\-ity \cite{gm}.  In particular, the strong
fluctuations of interparticle forces and the organization of the
largest of these in tenuous force networks have recently attracted
considerable attention 
\cite{network,network2,exp2,photoel,Pf,makse,qm}. The probability
density function of forces is thus a basic object of
study. Measurements \cite{network,network2,exp2,photoel}, numerical 
simulations \cite{Pf,makse} and theory \cite{qm} agree that such
force distributions decay exponentially for large forces \cite{deformnote}. 
The behavior for small 
forces is less well settled; while the $q$-model \cite{qm} seems to
predict a vanishing probability, experiments and numerical simulations
clearly show that this probability remains non-zero for small
forces. Since the small force distribution may be a fingerprint of
arching \cite{arch}, or of whether a system is jammed or unjammed
\cite{jamnote}, it is important to obtain a clear physical
interpretation of this discrepancy.

In this paper, we resolve this issue by elucidating the effect of the
{\em local} packing geometry on the force network in 2D packings of
frictionless, Hertzian spheres under gravity (see Fig.~1a).
%
%
To do so, it is important to distinguish the effective weight $W$ of a
particle $j$ from the interparticle forces $F$ (Fig.~1b). In the bulk,
this weight is carried by the other particles on which particle $j$
rests; however, the particles in the bottom layer are
typically supported by the bottom only, so the forces exerted on the
support are then to a good approximation the same as the weights. {\em Thus it
is essentially the distribution of weights ${\cal P}(W)$ which is
probed in experiments} where the particle-wall forces are extracted
from the imprints on carbon paper \cite{network,network2} or by force sensors
\cite{exp2}. Likewise, the main prediction of the $q$--model is for
the distribution ${\cal P}(W)$, rather than for the distribution of
interparticle forces, denoted by $P(F)$.

For our case of frictionless spheres, we define the weights $W$ as (see
Fig.~1b)
\begin{equation}
W_j \equiv m_j g + \sum_{<i>} (\vec{F}_{ij})_z~. \label{defw}
\end{equation}
Here $m_j$ denotes mass, $g$ denotes gravity, $\vec{F}_{ij}$ are the
interparticle forces, and the sum runs over all $n_c$ particles that
exert a force on particle $j$ {\em from above} (see Fig.~1b).  In the
following we rescale $W$ and $F$ to their average values, and write
the rescaled weights and forces as $w$ and $f$, with distributions
${\cal P}(w)$ and $P(f)$.

\begin{figure}[tb]
\includegraphics[width=8.0cm]{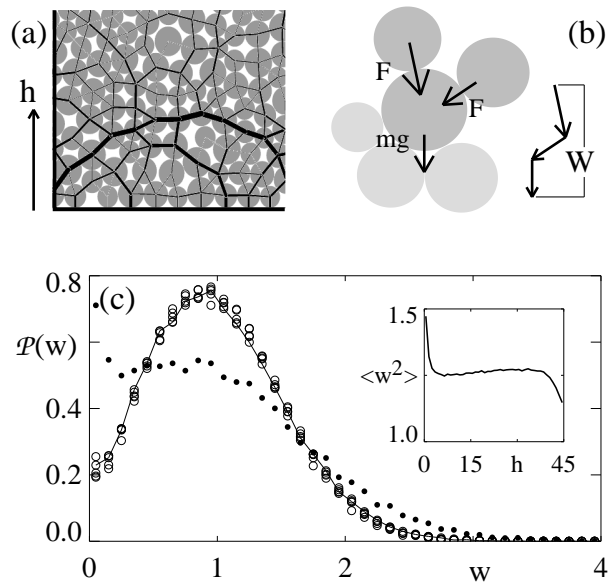}
\caption{(a) Detail of a typical packing and force network in our
simulations; the height $h$ denotes the distance from the bottom. 
(b) Definition of interparticle forces $F$ and weight $W$,
for a particle with $n_c \!=\!2$.  (c) The weight distribution ${\cal
P}({w})$ at various heights between 10 and 30 in the bulk (open
circles), for $2<h<3$ (full curve) and at the bottom (dots). Inset:
The second moment $\langle {w}^2 \rangle$ as a function of height $h$.
}\label{fig.rfbottom}
\end{figure} 

Our main findings are the following: {\em{(i)}} ${\cal P}(w)$ changes {\em
qualitatively} when approaching a boundary; in particular the
probability of finding a small weight is much larger at the bottom
than in the bulk (Fig.~1c).  {\em{(ii)}} The number of contacts from
above, $n_c$, crucially influences the distribution ${\cal P}(w)$, as can be
anticipated from Eq.~(\ref{defw}); the difference between bulk and
bottom ${\cal P}(w)$'s is almost entirely due to the change in $n_c$ caused
by the change in packing near the boundary (Fig.~2).  {\em{(iii)}} The
force probability distributions $P(f)$ and $P(f_z)$ show a much weaker
variation when approaching the boundaries (Fig.~3). {\em{(iv)}} The
distribution of $n_c$'s near the bottom can be manipulated by, e.g.,
curving the boundary of a highly monodisperse packing and this can
have a large effect on ${\cal P}(w)$ (Fig.~4).
 
\paragraph{Numerical method} Our 2D packings consist of frictionless
particles under gravity; the particles interact through normal
Hertzian forces, where $f\propto d^{3/2}$ and $d$ denotes the overlap
distance \cite{hertz}.  Unless noted otherwise, the material constants
and gravity are chosen such that a particle deforms $0.1\%$ under its
own weight, and the particle radii are drawn from a flat distribution
between $0.4 < r < 0.6$. Masses are proportional to the radii cubed. 
The container has a width of $24$, employs
periodic boundary conditions in the horizontal direction and has a
bottom consisting of a fixed hard support.  The data shown in this paper
were obtained from $1100$ realizations containing $1180$ particles
each. We construct our stationary packings by letting the particles
relax from a gas-like state by introducing a dissipative force that
acts whenever the overlap distance $d$ is nonzero.

\paragraph{Distribution of weights}
In Fig.~1c we show the weight distribution ${\cal P}({w})$ for the
bottom particles (dots) which differs qualitatively from the bulk
distributions (open circles). Moreover, we observe that the transition
is remarkably sharp: in the slice $2<h<3$, the weight distribution is
already bulk-like (full curve). In the inset of Fig.~1c we plot
$\langle {w}^2 \rangle$ which quantifies the width of ${\cal P}({w})$
as a function of height $h$. The sharp transition of ${\cal P}({w})$
near the bottom is clearly visible. Additionally, in the bulk,
$\langle {w}^2 \rangle$ slowly increases with height, due to the
deformations of the particles; this effect disappears for harder
particles \cite{prep}. Finally, near the top layer ${\cal P}(w)$
becomes sharply peaked and $\langle {w}^2 \rangle$ decreases.

To understand the change of ${\cal P}({w})$ near the bottom, consider
the typical packing of Fig.~2a. The support aligns the bottom row of
particles and thus strongly affects the directions of the
interparticle forces. The forces between neighboring bottom particles
are almost purely horizontal, so these hardly contribute to either the
force on the particle from above [the ``weight'' in (\ref{defw})] or
to the force needed to support it. This approximation becomes better the 
smaller the polydispersity is. Thus, the average
value of $n_c$, the number of particles that press on the particle
from above and hence contribute to its weight, is on average lower at
the bottom than in the bulk. Intuitively it is clear that {\em the
probability of finding a small value of $w$ increases with smaller
$n_c$} for non-tensile forces. This statement can be made precise by
considering Eq.~(\ref{defw}) for fixed $n_c$ and analyzing ${\cal
P}_{n_c}(w)$, the weight probabilities restricted to particles of
given $n_c$.  As long as the joint probability distribution of the
interparticle forces remains finite for small forces,
it follows from a phase-space argument that
\begin{equation}\label{smallfz}
{\cal P}_{n_c}({w})\propto {w}^{{n_c}-1} \quad\quad {\rm for} \quad w
\rightarrow 0~,
\end{equation}
for all $n_c \! \geq\! 1$. The particles which do not feel a force
from above, $n_c\!=\!0$, give a
$\delta$--like contribution at $W\!=\!1$;  for deep layers this occurs
for ${w}\!  \ll \!1$.

\begin{figure}[t]
\includegraphics[width=8.0cm]{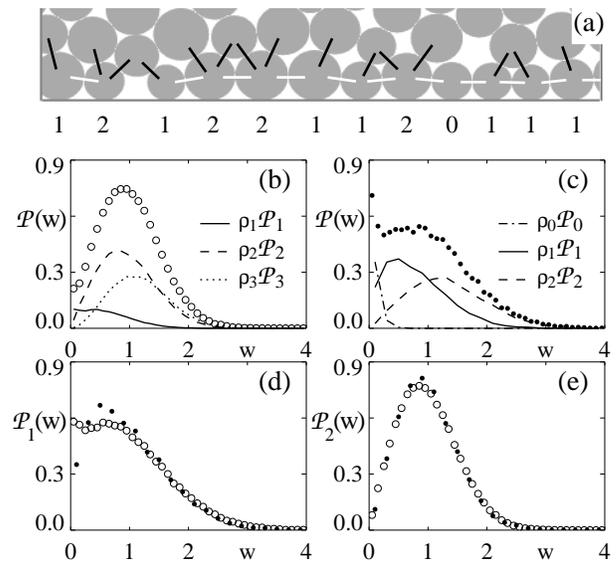}
\caption{(a) Detail of a typical packing, showing the dominance, near
the bottom, of layer-to-layer forces (black lines) to intralayer forces
(white lines) in determining $w$. The numbers show the values of
$n_c$ for the respective bottom particles.  (b,c) Decomposition of
${\cal P}(w)$ according to Eq.~(\ref{rtotfz}) in the bulk (b) and at the
bottom (c); The measured bulk values for the fractions
$\{\rho_0,\rho_1,\rho_2,\rho_3\}$ in Eq.~(3) are
$\{0.01,0.11,0.52,0.36\}$, and the bottom values are
$\{0.08,0.46,0.44,0.02\}$; at the bottom, we excluded the intralayer,
almost horizontal forces.  (d,e) When rescaled to the average value
for each distribution function, ${\cal P}_1(w)$ (d) and ${\cal
P}_2(w)$ (e) are essentially the same in the bulk (open circles) and
at the bottom (dots).}
\label{fig.decomp}
\end{figure}

To check the validity of this idea,  we have
determined ${\cal P}_{n_c}(w)$ both in the bulk and near the bottom by
determining  $n_c$ for each particle and
decomposing the weight distribution ${\cal P}({w})$  into the 
${\cal P}_{n_c}({w})$'s,
\begin{equation}\label{rtotfz}
{\cal P}({w})=\sum_{n_c} \rho_{n_c} {\cal P}_{n_c}({w})~.
\end{equation} 
The crucial difference between bottom and bulk are the {\em
fractions} $\rho_{n_c}$ of particles that feel $n_c$ other particles
pressing on them from above. This can be seen in
Figs.~\ref{fig.decomp}b-c, where we show the decomposition of ${\cal
P}(w)$ according to Eq.~(\ref{rtotfz}). This picture is also confirmed
by Figs.~\ref{fig.decomp}d-e, which show that the individual
distribution functions ${\cal P}_{1}(w)$ and ${\cal P}_2(w)$ do not
differ significantly between bulk and bottom (to compare these, we
have to normalize them not to the total average weight in each layer,
but to the average weight of the particles with the same $n_c$).  Note
also that Figs.~\ref{fig.decomp}b-e show that
Eq.~(\ref{smallfz}) is valid, {\em except for ${\cal P}_1({w})$ at the
bottom for small values of $w$}; this deviation comes from neglecting
the intralayer, ``almost horizontal'' forces, whose small vertical
components eventually make ${\cal P}_1({w}) \rightarrow 0$ for very
small weights.

\begin{figure}[t]
\includegraphics[width=8.0cm]{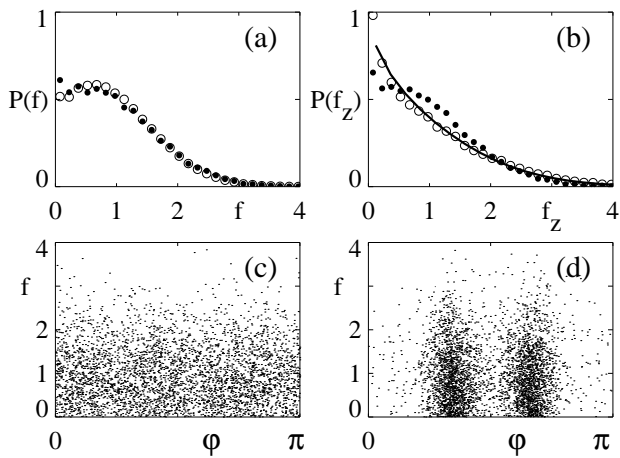}
\caption{(a) $P(f)$ in the bulk (open circles) and for the
layer--to--layer forces (see Fig.~2a) near the bottom (dots).  (b)
$P(f_z)$ in the bulk (open circles) and for the layer--to--layer
forces near the bottom (dots). The solid line is obtained by
integrating $P(f)$ over all angles (see text). (c,d) Scatter plot of
$(f_{ij},\varphi_{ij})$ for (c) the bulk forces and (d) the
layer--to--layer forces near the bottom. Same packings as for 
Figs.~\ref{fig.rfbottom} and~\ref{fig.decomp}.}
\label{fig.pv}
\end{figure}

\paragraph{Interparticle forces}
The distributions of the interparticle forces are much more robust
than the weight distributions. The results for the distribution
$P(f)$ of $|\vec{f}|$ are shown in Fig.~\ref{fig.pv}a. The only
difference between bulk and bottom distributions is that the small
{\it peak} around $f=0.7$ for bulk forces becomes a {\it plateau} for
$P(f)$ for the forces close to the bottom (see Fig.~2a). It is
intriguing to note that this change from a plateau to a peak is
reminiscent of what is proposed as an identification of the {\it
jamming} transition \cite{jamnote}.

We also have measured the distribution of angles $\varphi_{ij}$, which
define the orientation of the $\vec{f}_{ij}$, and find that these
angles are uniformly distributed and independent of the absolute value
of $\vec{f}$ in the bulk, see Fig.~\ref{fig.pv}c. Thus, in the bulk
our packing is isotropic. Near the boundary, however, this isotropy is
broken strongly: in agreement with our scenario for the influence of
the packing geometry, the angles of the forces between bottom particles 
and those in the layer above 
are concentrated around $\pi/3$ and $2 \pi/3$, as Figs.~\ref{fig.decomp}a 
and~\ref{fig.pv}d show.  Near the bottom, therefore, the interparticle 
forces naturally divide up into almost horizontal intralayer forces and 
``layer-to-layer'' forces.

Since the weight distribution is determined by the $z$-components of
the forces, let us also investigate the distribution $P(f_z)$.
According to Fig.~\ref{fig.pv}b, $P(f_z)$ also remains non-zero for
small $f_z$, both in the bulk and near the bottom.  There is a
substantial difference, however, associated with the difference in
packing. In the bulk, we have seen in Fig.~\ref{fig.pv}c that there is
no noticeable correlation between the angles $\varphi_{ij}$ and the
force strength $f_{ij}$. Hence in the bulk $P(f_z) \approx \int d
\varphi d f ~ P(f) P(\varphi)~ \delta(f_z- f sin (\varphi)) $ with
$P(\varphi)= const$.  Indeed, the distribution obtained by numerical
integration of this relation with $P(f)$ from Fig.~\ref{fig.pv}a and a
uniform angle distribution yields the solid line in
Fig. \ref{fig.pv}b, which closely follows $P(f_z)$ as measured in the
bulk.  Near the bottom, on the other hand,
the value of $\sin(\varphi)$ is concentrated around
${1\over2}\sqrt{3}\approx0.866$: in the approximation that the
distribution of $\sin(\varphi)$ is sharply peaked at this value, the
shape of $P(f_z)$ is close to that of $P(f)$, which is confirmed by
direct comparison of the dotted datasets of panels a and b of
Fig.~\ref{fig.pv}.

\begin{figure}[floatfix]
\includegraphics[width=8.0cm]{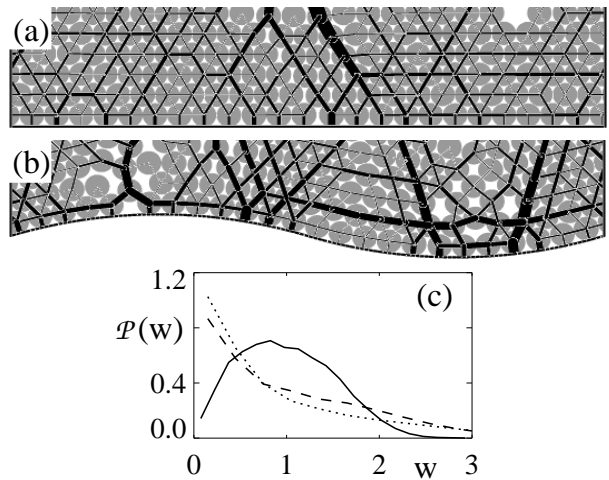}
\caption{(a,b) Packing and force networks in a weakly polydisperse
packing near a flat bottom (a) and a curved bottom (b). (c)
Distribution of weights ${\cal P}(w)$ on the flat bottom (solid line), convex
curved bottom (dashed line) and the concave curved bottom (dotted
line). The various shapes originate from the corresponding
$\{\rho_0,\rho_1,\rho_2\}$: $\{0.00,0.10,0.90\}$, $\{0.02,0.39,0.58\}$ and 
$\{0.04,0.46,0.50\}$ respectively. }\label{fig.monocurved}
\end{figure}

\paragraph{Manipulating ${\cal P}(w)$}
In the previous paragraphs we have shown that the weight distribution
${\cal P}(w)$ is very sensitive to the local packing geometry, while
the distribution of interparticle forces is robust. This allows one to
manipulate ${\cal P}(w)$ at the bottom by changing the boundary
conditions.  We illustrate this by simulations of weakly polydisperse
particles, $0.49<r<0.51$, both with a flat bottom
(Fig.~\ref{fig.monocurved}a) and a {\it curved} bottom, consisting of
two circle segments of radius 20 glued together
(Fig.~\ref{fig.monocurved}b). For the flat bottom, the particles form
an almost perfect hexagonal packing, leading to particles with mostly
$n_c=2$ (horizontal contacts again not included). In agreement with
(\ref{smallfz}), ${\cal P}(w)$ increases linearly for small $w$ in
this case. The weakly curved bottom locally disturbs this crystalline
structure (Fig.~4b), causing a dramatic change in the fractions
$\rho_{n_c}$, and correspondingly in ${\cal P}(w)$
(Fig.~\ref{fig.monocurved}c). Note that the small difference between
the distributions of the convex and the concave part of the bottom are
reflected in the $\rho_n$ as well.  Interestingly, $P(f)$ is in all
these cases indistinguishable from $P(f)$ in the strongly polydisperse case
\cite{prep}.

\paragraph{Perspective} The message that emerges naturally from the
above analysis is clear: in experiments in which the forces on a boundary 
are probed, one measures effectively the weights $w$. These weights,
however, are not the most fundamental quantities of a granular
packing, as they are derived from the {\it interparticle forces}
$\vec{f}_{ij}$. These capture the full microscopic structure, and the
distribution function $P(f)$ of the force {\em strength} is quite
insensitive to the packing, in contrast to ${\cal P}(w)$. Simple phase
space considerations show that a grain with $n_c$ contacts with
particles that press downwards on it makes a contribution to the weight
distribution ${\cal P}(w)$ which scales as $w^{n_c-1}$ as $w \to
0$. Thus, the small weight distribution is dominated by particles with
few such contacts, in particular by those with just one, and the
change in geometry near boundaries leads to an atypical ${\cal P}(w)$.  
In addition, we occasionally observe a
small peak at $w\ll1$ due to ``loose'' particles ($n_c\!=\!0$) which
do not feel a force from above; there are indications for such a peak
of ${\cal P}(w)$ at $w=0$ in recent precise experiments (see Fig.~5 of
\cite{network2}).


In the standard $q$-model, weights are randomly distributed over a
fixed number of neighbors one layer below; in the simplest version
there are two such neighbors that receive a fraction $q$ and $1-q$ of
the weight \cite{qm}. Due to a fixed connectivity, this model cannot
be expected to capture the behavior of ${\cal P}(w)$, especially near
boundaries.  The {\em product} of the weight $w$ and $q$, however,
could be interpreted as an interparticle force. Interestingly, in the
simplest case of a uniform probability distribution of the $q$'s, the
probability distribution $P( q w) $ is a pure exponential
\cite{prep,snoeijer}. In simulations of the $q$-model with random
connectivities, a variety of ${\cal P}(w)$'s can be obtained, similar 
to what we found here for the frictionless spheres \cite{prep}.

It is possible to test our framework in experiments.  For
${\cal P}(w)$ we expect that curvature effects, such as shown in
Fig.~4, only play a role when they break highly ordered packings;
indeed carbon-paper measurements at the sidewalls in cylinders give a
very similar ${\cal P}(w)$ as near the bottom \cite{network}. A
simpler way to change the boundary conditions may be to include a
layer of larger particles at the boundary; their value of
$n_c$ will be higher, and we expect that ${\cal P}(w)$ for small $w$
will decrease for larger $n_c$. Furthermore, there is a strong need of
direct determination of $P(f)$, both in the bulk and near the
boundaries, since the present data concerns ${\cal P}(w)$ only
\cite{network,network2,exp2}.  An interesting observation in the
context of ``jamming'' \cite{jamnote} is that in our simulations the 
peak in $P(f)$ appears to vanish near the boundaries (Fig.~3a); are 
granular materials no longer jammed here, and is this relevant for the
localization of shear bands near boundaries?

An important issue for future study is clearly the role of friction
and dimensionality. Our numerical study has been done in two
dimensions with frictionless spheres; however, recent studies indicate
\cite{makse} that the coordination number for 3D packings with
friction is similar to those of 2D frictionless
packings. Qualitatively, the picture we have advanced is therefore
expected to capture the realistic case of three dimensions with
friction, because our phase space arguments are independent of
dimension.

We finally note an interesting open issue. In the $q$-model, ${\cal
P}(w)$ approaches its asymptotic expression as a function of depth
algebraically slow \cite{snoeijer}). In our packings, the convergence
appears to be much faster --- is this a real discrepancy and if so is
it related with the same packing issues?

We are grateful to Martin Howard and Hans van Leeuwen for numerous 
illuminating discussions.


\begin{thebibliography}{99}
\bibitem{deformnote} This tail is only exponential provided that the
deformations of the particles are sufficiently small \cite{makse}.
\end{thebibliography}
\end{document}